# Distinguishing strain, charge and molecular orbital induced effects on the electronic structure: graphene/ammonia system

Tanmay Das,[a)] and Sesha Vempati

*Department of Physics, Indian Institute of Technology Bhilai, GEC Campus, Raipur-492015.*

Molecular adsorption at the surface of a 2D material poses numerous questions regarding the modification to the band structure and interfacial states, which of course deserve full attention. In line with this, first principle density functional theory is employed on graphene/ammonia system. We identify the effects on the band structure due to strain, charge transfer and presence of molecular orbitals (MOs) of $NH_3$ for six adsorption configurations. Induced-strain upon ammonia-adsorption opens the band gap ($E_g$) of graphene due to the breaking of translational symmetry and shifts the equilibrium Fermi energy ($E_F$). The $E_g$ and $E_F$ values and charge density distribution are dependent on the adsorption configuration, where the MO structure of $NH_3$ plays a crucial role. The presence of MOs of N or H -originated pushes the unoccupied states of graphene towards $E_F$. $NH_3$ forms an interfacial occupied state originating from $N2p$ below the $E_F$ within ~1.6 – 2.2 eV for all configurations. These findings enhance fundamental understanding of graphene/$NH_3$ system.

___________________________

[a)] Author to whom correspondence should be addressed. Electronic mail: tanmayd@iitbhilai.ac.in

**Introduction**

Recently two-dimensional (2D) materials have emerged as excellent candidates for the detection of individual gas molecules.[1, 2] Among various materials, there are a number of reasons that make graphene a promising material towards detection of single molecules.[3] For instance, the lower density of scattering centers and ballistic transport, functional dependence of conductivity on the carrier concentration etc are notable features apart from high surface area to volume ratio originating from the 2D structure. In general, the origin of the gas-sensitivity is attributed to a possible electronic interaction of graphene with the test molecule, that causes a detectable change in the conductivity.[4] Various molecules in gaseous state like $H_2O$, $NH_3$, CO, $NO_2$, NO have been subjected to investigation with first principle study [5, 6] as well as in experiment.[6] Indeed, as an additional advantage, some of these molecules were found to be useful as a molecular dopant which open the band gap [7, 8] of graphene with some degree of tunability.[9, 10] In any case, deeper understanding of the cause of band gap renormalization and effect on the conductivity is of utmost importance. Experimentally, the adsorption and desorption of single ammonia molecule on graphene has shown the quantization of changes in graphene conductivity by signaling the adsorption or desorption of each individual molecule, which supported the donor nature of $NH_3$.[3] However, the first principle studies [5, 6] have indicated relatively weak adsorption energy and small [11] or almost no charge transfer from $NH_3$ to graphene. Interestingly, the magnitude of the charge transfer is expected to be dependent on the orientation of the molecular orbitals (MOs) of $NH_3$. If the hydrogen atoms are positioned towards (down, D) or away (up, U) from graphene a charge transfer of zero or $\sim 0.048 \times 10^{-19}$ C/molecule is observed, respectively. When the test molecule arrives at the surface the interaction (Coulombic and/or Van der Waal's) reorganizes the potential energy landscape inducing some strain on graphene apart from a possible charge transfer. Therefore, the change in the conductivity may be due to renormalized band structure which may have contribution from either (i) strain, (ii) charge transfer and/or (iii) spatial arrangement of MOs of the host that may act as scattering centers. Theoretical studies indicated modifications to the band structure due to external *strain*.[12] It is also found that depending on the strength and the direction of strain, band gap can be tuned in the order of 0.1 eV. Similarly, band structure may be modified due to the transfer of charge.[13, 14] The net strain is a result of interaction between substrate/graphene and graphene/molecule, while the charge transfer can be dependent on the orientation and the symmetry of the molecule. Indeed, the previous studies [5, 6] assumed no significant changes to the band structure upon adsorption of $NH_3$ and did not distinguishing the strain and charge transfer-induced changes. Such an understanding is rather beneficial to unfold the mechanism of molecular doping and applicability of graphene in electronic devices and sensors, where the selectivity towards a particular molecule is challenging.[15, 16]

In this letter, we have investigated the effects of adsorption of ammonia on the band structure of graphene from two different perspectives *viz* strain and MOs/charge -induced. Importantly, we found that the adsorption of $NH_3$ opened a direct small energy gap of about ~200 meV at high symmetry point K. In our calculations we have observed explicit differences to the occupied and unoccupied states which depends on the adsorption configuration with respect to the symmetry of $NH_3$ molecule. In addition to this, depending on the MO orientation, $NH_3$ forms an occupied state originated from N2$p$ below the $E_F$ at 1.6 – 2.2 eV.

**Computation**

Adsorption of ammonia molecule on pristine graphene was examined by first-principle methods based on Density Functional Theory, implemented in the Quantum Espresso package.[17] The exchange-correlation functional of Perdew–Burke–Ernzerhof (PBE) approximated by Generalized Gradient Approximation was used for the calculation.[18] Norm-conserving ultrasoft pseudopotentials (Rappe Rabe Kaxiras Joannopoulos) were used to describe the interaction between ionic cores and valence electrons.

A $3 \times 3$ supercell of graphene containing 18 carbon atoms was used to minimize the interaction between the molecule and their periodic images. The z-axis of the periodic supercell (i.e., normal to the graphene surface) was taken as much as 16 Å to suppress the interaction between the graphene sheet and adsorbed molecule of the adjacent supercell. In our calculation, structural optimization was performed by relaxing a single molecule of ammonia on graphene by intuitively placing it at high symmetry positions (C, B, and T sites) (FIG 1) with a convergence threshold of $10^{-3}$ Ry/Bohr. This process is repeated for both U and D configurations. We refer C position with U configuration as CU in short. Similar nomenclature applies to BU, BD, TU and TD configurations. A plane-wave basis set with a cutoff energy of 800 eV was employed and non-spin-polarized calculations were used for all the configurations. The Brillouin zone integration involved in calculating the system's electronic density of states (DOS) was performed with a $3 \times 3 \times 3$ Monkhorst-Pack [19] grid with a Gaussian broadening of 0.05 Ry. The band structure calculations were done by taking the wavevector path as *K-Γ-M-K*. After optimization of the geometry of graphene/$NH_3$ system, calculation was performed for two cases. In case 1, the very same optimized graphene/$NH_3$ system was considered and in case 2, $NH_3$ was removed (named as Gr-strain) however the coordinates of the C-atoms were frozen. The results are compared with relaxed counterparts. Graphene/$NH_3$ system is denoted as Gr/$NH_3$ which accounts for charge transfer and strain induced effects. Gr-strain considers the deformation of graphene (i.e. accounts for strain induced effects only) due to $NH_3$ adsorption. Note that the induced strain is a function of adsorption configuration and adsorption site.

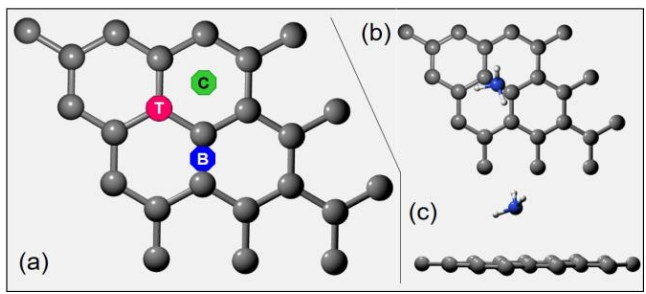

**FIG 1: (Color online) Schematic of graphene/$NH_3$ (a) optimized geometry and annotation of spatial positions of high symmetry positions (C, B, and T sites), while (b) and (c) shows top and cross-sectional view of BU configuration, respectively.**

**Results and discussion**

In order to distinguish the effects from strain and that of charge transfer on the electronic structure of a hybrid system, it is inevitable to study their isolated counterparts. To start with, isolated ammonia is a closed shell molecule with an intrinsic dipole moment of $3.3 \times 10^{-30}$ Cm that arises due to the asymmetric charge distribution on the nitrogen atom.[20] $NH_3$ is known to interact with graphene weakly through van der Waals forces where frontier molecular orbitals (MO's) play a crucial role. Hence we have calculated the DOS and the frontier MO's of $NH_3$ and the results are shown graphically in FIG 2a. The electronic geometry of $NH_3$ is tetragonal due to the unshared pair of electrons on the N atom. From the total DOS of $NH_3$ (FIG 2c) we have identified highly occupied molecular orbital (HOMO) and least unoccupied molecular orbital (LUMO).[5] Our calculation indicates that the HOMO of $NH_3$ is located at ~13.6 eV, which is comparable with the experimental ionization potential of 10.2 eV.[21] For the case of pristine graphene, we have observed (results not shown here) a uniform accumulation of charge around the carbon atoms, along the bonds, however, at the center of the honeycomb the charge distribution is found to be zero. This minimum charge accumulation site is understandable, in the light of the electronegativity values of C(2.55) and H(2.1). Due to the strong covalent bonding between C-C atoms maximum charge is accumulated uniformly around C atoms and along the bonds.

For graphene/$NH_3$ system, the charge accumulation is significantly different from their pristine counterparts, although they are characterized as weakly interacting. In this context, with reference to the phase of the electron wavefunction on graphene, FIG 2b shows the localization of LUMO* and HOMO* under equilibrium condition.[22] Hence the spatial distribution of charge density is a result of *net* charge transfer, which depends on the adsorption configuration.[5] On the other hand, after structural optimization of graphene/$NH_3$ system carbon atoms are displaced in the order of 0.2 Å which is relatively more in case of U orientation than that of D. The distance between C and N atom after structural relaxation is shown in FIG 2d (right) for CD and CU configurations. The relative atomic positions of nitrogen and hydrogen in $NH_3$ vary slightly after adsorption on graphene, however, this slight displacement, did not cause any significant changes to the DOS of $NH_3$. This is in contrast to that of graphene and will be discussed later. Furthermore, in the context of charge transfer, the relative alignment of the Fermi level and HOMO-LUMO of the isolated systems enable the transfer, see FIG2d (left). Essentially, the energetic position of $E_F$ of graphene within the gap of the HOMO and LUMO of an isolated molecule is the key for the donor or acceptor nature.[23] In FIG 2d (right) we have shown the equilibrium distance between carbon and nitrogen atoms for CD and CU configurations. In the following, with reference to FIG 3, we further discuss the effect of this relative positioning of HOMO and LUMO.

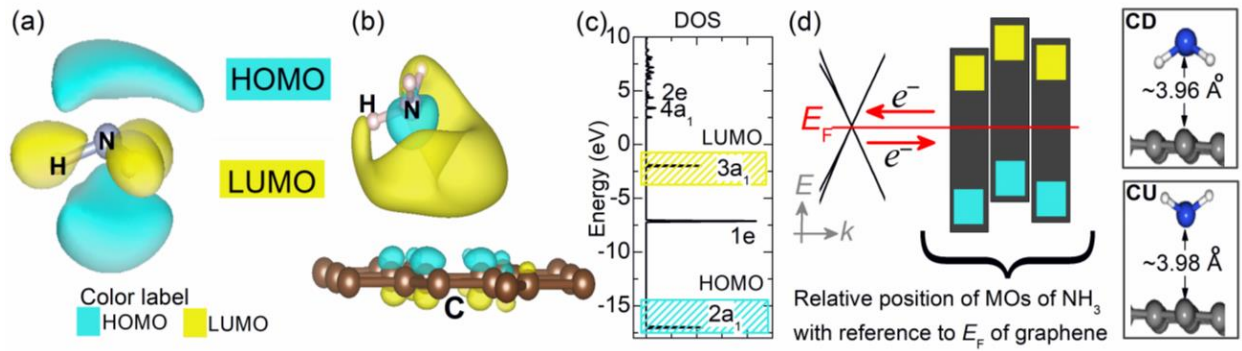

**FIG 2:** (Color online) Frontier molecular orbitals (MOs) of (a) isolated NH$_3$ (b) graphene/NH$_3$ in CU configuration and (c) density of states (DOS) of isolate NH$_3$. (d) (left) schematic of energy band diagram for graphene/NH$_3$ depicting relative positions of the HOMO-LUMO gap with reference to $E_F$ of graphene and (right) shows equilibrium distance between N and C atoms for CD and CU cases.

The electronic coupling between NH$_3$ and graphene would depict multifold effects on the electronic structure, specifically on the $E_F$ and band gap ($E_g$) values, apart from changes to the conductivity of the substrate as observed earlier.[3, 13, 14] We have also determined the $E_F$ and $E_g$ values in the presence and absence of NH$_3$. The deformation (strain) of graphene lattice in the presence of NH$_3$, *without* the molecule yields a different $E_F$ value ($E_F^{Strain}$) when compared to the case if the molecule is present ($E_F^{Gr/NH_3}$). The corresponding band gap values are denoted as $E_g^{Strain}$ and $E_g^{Gr/NH_3}$, respectively. The $E_F$ values for isolated graphene ($E_F^{Gr}$) and NH$_3$ are calculated and found to be 1.89 eV, and 3.57 eV, respectively. FIG 3 (bottom) shows the shift, ($E_F^{Gr} - E_F^{Strain}$) and ($E_F^{Gr} - E_F^{Gr/NH_3}$) for the six configurations with reference to $E_F^{Gr}$. There is almost no change to the $E_F$ values in the case of TD, BD and CD due to strain. Upon adsorption of NH$_3$, the $E_F^{Gr/NH_3}$ values are increased, with small variations among TD, BD and CD. In contrast to D-, for U- configuration, the strain-induced a reduction in the $E_F$ values by ~20-30 meV. More interestingly, upon adsorption of $NH_3$, $E_F^{Gr/NH_3}$ values have shown a further reduction by ~500 meV, where TU-type adsorption is at the lowest ($E_F^{Gr/NH_3}$ = 1.37 eV).

$E_g^{Strain}$ and $E_g^{Gr/NH_3}$ are plotted in FIG 3 (top) for the six configurations, which range from ~180 – 230 meV. Strain-induced opening of band gap of graphene is attributed to the breaking of the translational symmetry.[24] The presence of $NH_3$ nominally altered the band gap values for all configurations. This is rather interesting finding, where the presence of MOs and/or charge induces modification to the $E_g$. Note that the occupied molecular orbitals (HOMO*) near to the graphene-surface would experience Coulombic and Pauli repulsion with the electrons in the VB of graphene.

The electronic interaction between substrate and adsorbate can be characterized based on the coupling between molecular orbitals and the substrate electronic states. i.e. alignment of energy levels and localization of electrons.[25] The combined effect of these main factors results in either integral or fractional transfer of charge, apart from any Coulombic and screening effects.[26] For instance, as found in the case of organic/metal interface, the alignment of the energy levels at the interface related to the charge transfer between the donor or acceptor species and the metallic surface.[27] It is also found that the charge transfer alters the alignment of energy levels and causes structural transformations in both donating and accepting species.[28] Essentially, the potential energy landscape would be re-normalized to accommodate the extra or deficit electronic charge which leads to the (local) structural deformation. This deformation causes stress on both substrate and adsorbate while the magnitude of which depends on the microscopic mechanical properties. Indeed, our approach distinguishes the effects on $E_F$ and $E_g$ due to the presence of (i) MOs and/or charge of $NH_3$ from that of (ii) deformation of the graphene basal plane.

For graphene/$NH_3$ case, the first principle studies [5, 6] have indicated relatively weak adsorption energy and small [11] or almost no charge transfer from $NH_3$ to graphene. On the other hand, the donor nature of $NH_3$ is experimentally supported.[3] Indeed, D and U type adsorptions are fundamentally different, where the atomic orbitals of hydrogen and nitrogen are closer to the graphene basal plane, respectively (FIG 2b). Consequently, the charge-transfer banks on the orientation of the MOs of $NH_3$. If the hydrogen atoms are positioned towards (D) or away (U) from graphene a charge transfer of zero or ~$0.048 \times 10^{-19}$ C/molecule is observed, respectively. Since the loan pair of electrons in $NH_3$ are localized on the nitrogen atom (FIG 2a), D-type adsorption may not be an ideal configuration for transfer of charge, where we don't expect any wavefunction overlap. In contrast to this, U-type adsorption is relatively favorable for the transfer of charge.

Given this background, it is convincing that the configuration-dependent changes to $E_F$ and $E_g$ lean on various factors such as the strength and type of interaction between the adsorbent and the substrate.[7-10] Depending on the degree and type of deformation, we have observed some changes to the $E_F$ values for graphene/$NH_3$. In the case of graphene, it is known that the $E_F$ is rather sensitive to the doping density, which might happen due to the presence of $NH_3$. Also, one may expect changes to the $E_F$ when the translational symmetry is broken due to strain.[7-10, 12] For D configuration the effect from MOs of $NH_3$ are predominant, while for U, we expect effects from MOs as well as charge (FIG 2b) on $E_F$ and $E_g$ values.

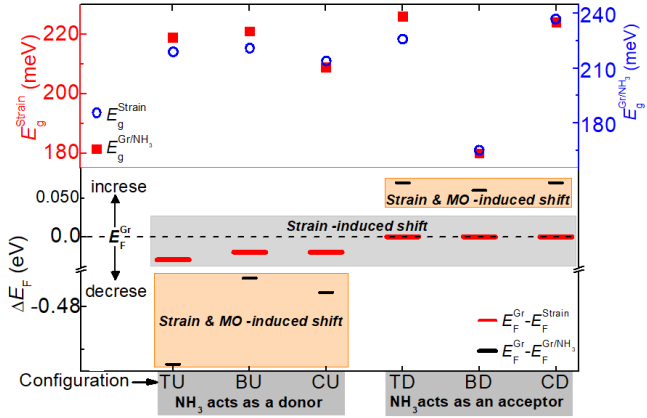

**FIG 3: (Color online) Energy band gap ($E_g$) and shift of Fermi Energy ($E_F$) of graphene/NH$_3$ system with reference to that of pristine graphene are compared for six different adsorption configurations.**

As discussed earlier, slight variation in the atomic positions of graphene (strain), any transfer of charge and non-uniform charge distribution/HOMO, LUMO localization indeed influences the (un)occupied bands for both U and D cases. In order to understand the strain induced changes we have computed the band structure of relaxed graphene and discussed in the following. PDOS of graphene are shown in FIG 4a and the corresponding band structure is shown in FIG 4c. First Brillouin zone for graphene is shown as inset of FIG 4a. Band structures were computed for all the six combinations and the results are shown in SFIG 1. Indeed, apart from some minor changes, the band structures for all U configurations (CU, BU and TU) are rather similar, however, different from that of all D counterparts (CD, BD and TD). See SFIG 1 and the interpretation therein. Hence we have selected BU and BD configurations for the discussion where in U(D) configuration, NH$_3$ not only acts as donor [5] (acceptor) but also induces a finite strain on the graphene lattice.

In the context of relaxed graphene, carbon atoms are arranged in a regular honeycomb structure due to their sp$^2$ hybridization of 2s, 2px and 2py orbitals accommodating three out of four valence electrons. The planer sp$^2$ hybridization in graphene is due to the mixing of the 2s and two 2p orbitals. The electrons involved in sp$^2$ hybridization are localized and form σ-bonds, which do not contribute to the transport phenomenon. Again, the π-bond with 2pz electrons of neighboring carbon atoms are formed by the fourth electron, perpendicular to the plane of graphene sheet, which occupies the 2pz orbital.[29] The electrons forming π-bond are delocalized over the entire lattice and have higher energy than the electrons that form the σ-bonds. These delocalized π-electrons are responsible for most of the extraordinary properties of graphene.[30] π* and π bands correspond to the bands above and below the Fermi level respectively. If we consider pure graphene, these π bands meets at the Fermi level with zero band gap (FIG 4c), which is the so called Dirac Point, and the region around them is known as Dirac cone.

Moving onto the strain-related effects, the changes are not just limited to the Dirac cone with an induced band gap, rather they extend into CB and VB of graphene. Figure 4b and d depict the band structure due to deformed graphene lattice in the absence of NH$_3$ for BD and BU configurations, respectively. For D-type adsorption, the energy levels associated with the σ$_{CC}$ bands (VB region, ~3 eV below $E_F$) split at *K* position due to the strain. For U-configuration, this split is not observed and the σ$_{CC}$ bands are isoenergetic with that of pristine graphene. On the other hand, for both D and U cases, we have observed some CB replicas due to the strain (annotated with 'R' on FIG 4b and d). To place in the context, the displacement of carbon atoms of ~0.22 Å from their equilibrium position.

With reference to the MOs and charge induced changes to the band structure, the strain and charge/MO-induced band gap is discussed earlier. We have identified some changes to the anti-bonding, $\sigma^*$ states. Previously mentioned CB-replicas have shifted towards $E_F$ due to the presence of $NH_3$. Interestingly, the magnitude of shift from their corresponding CB-replicas is more for U than that of D type adsorption configuration (SFIG 2) due to the possible transfer of charge in U-configuration. Transfer of charge, dopes the system which in turn reorganizes the potential energy landscape and adjustment of density of states to accommodate the changes to the charge density. Most importantly, in D and U configurations, N2p majorly contributes to a non-dispersive occupied band at -1.64 eV and -2.18 eV, respectively. We point out some minor contribution from N1s to this occupied state. Also, in BD configuration, the PDOS of N2p state at -1.64 eV is relatively broader than that of BU counterpart at -2.19 eV indicating relatively more electron delocalization. The non-dispersive nature of the occupied state from graphene/$NH_3$ might be arising due to the localization of the MOs on $NH_3$ molecule and interfacial in nature. Indeed, an occupied state just below the $E_F$ due to nitrogen is observed when p-quinquephenyl pyridine molecule is adsorbed at the surface of ZnO.[31] The presence of MOs of $NH_3$ and/or charge has no noticeable effect on the occupied states of graphene. Indeed PDOS contribution indicated almost no contribution of unoccupied states associated with $NH_3$. N1s also has some unoccupied states which however, differ in energy from D and U configurations (FIG 4a and e, $DOS^{N1s} \times 15$).

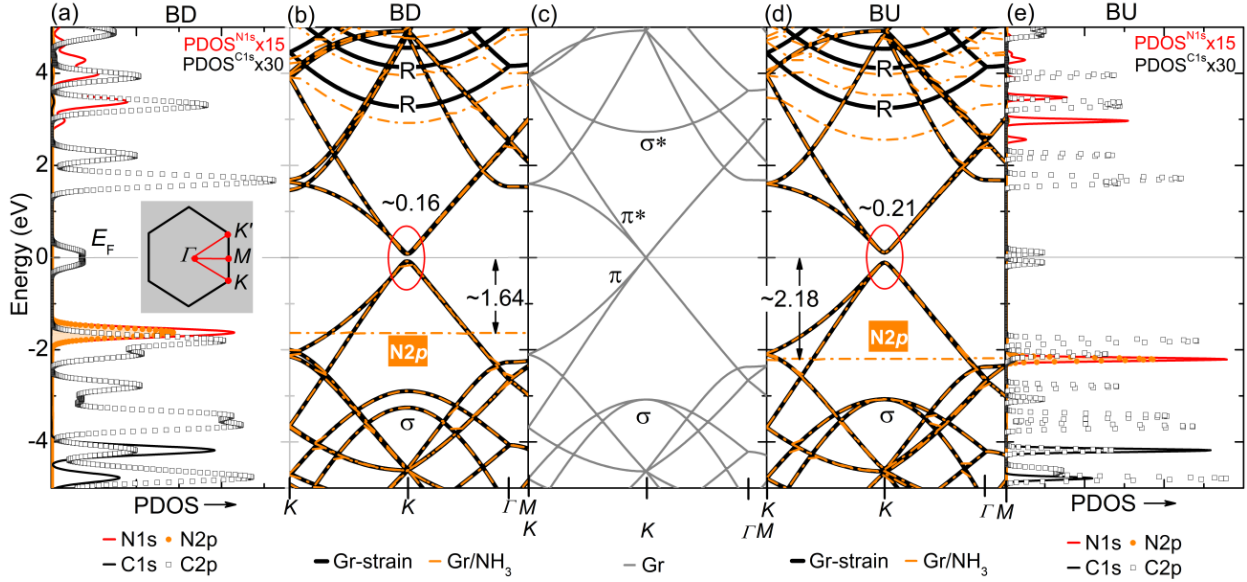

**FIG 4: (Color online)** Partial density of states (PDOS) and band structures for BD and BU configurations are juxtaposed with that of pure graphene. (a) PDOS of BD (b) band structure of BD for two cases graphene under strain (Gr-strain) and Gr/NH$_3$, (c) band structure of pure graphene, (d) band structure of BU for Gr-strain and Gr/NH$_3$ (e) PDOS of BU. Inset of Part (a) shows first Brillouin zone with the high-symmetry points $\Gamma$, $M$, and $K$. DOS$^{C1s}$×30: intensity of C1s PDOS is multiplied 30 times. Similar nomenclature applies to DOS $^{C1s}$×15 etc. Energy differences are shown in the units of electron volts.

**Conclusions**

Our approach on the graphene/NH$_3$ system unfolded effects of strain and charge related effects including spatial localization of MOs of NH$_3$. Isolated graphene depicted a uniform charge distribution around carbon atom and almost none at the center of the hexagon. For isolated NH$_3$, being in tetragonal structure, the electron density is localized on N atom. Also, we find that charge density distribution of isolated ammonia is quite distinct from that when adsorbed at the surface of graphene. We also find that charge density distribution of graphene/

$NH_3$ is quite distinct from that of isolated counterparts. Due to the re-organization of potential energy landscape at the surface of the graphene, the presence of $NH_3$ induced some strain on the graphene. Adsorption induced strain break the translational symmetry of graphene thus opening the band gap of graphene about 200 meV, which however, depends whether the hydrogen atoms of $NH_3$ point toward or away from graphene. The presence of MOs of $NH_3$ pushes the unoccupied states of graphene towards the $E_F$. For U and D configurations, the unoccupied states close to Fermi energy depict relatively larger shifts. HOMO* localization occurred within the interface between graphene and $NH_3$. We conclude that the effects on band structure of graphene are dependent on the symmetry and spatial distribution of MOs of $NH_3$ and type of adsorption.

# Distinguishing strain, charge and molecular orbital induced effects on the electronic structure: graphene/ammonia system

Tanmay Das* and Sesha Vempati

*Department of Physics, Indian Institute of Technology Bhilai, GEC Campus, Raipur-492015*
*tanmayd@iitbhilai.ac.in*

**Supplementary material**

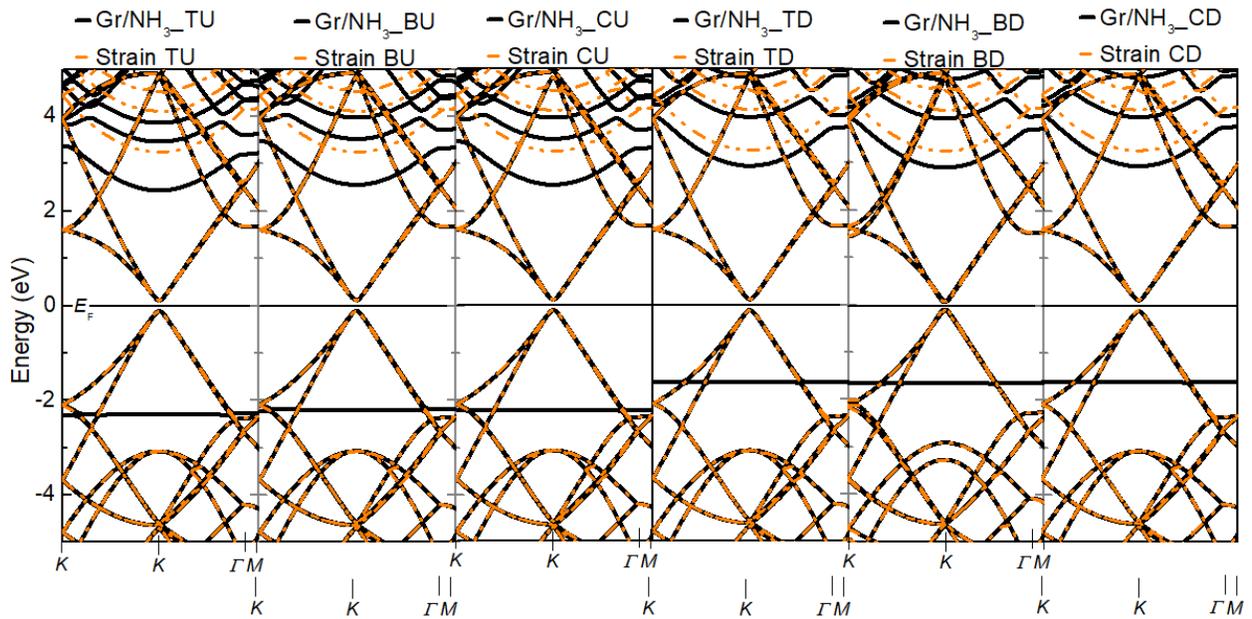

**SFIG 1: Electronic band structure of graphene/NH$_3$ system, where NH$_3$ is in six different orientations.**

Discussion: We observe that all bands overlap for both Gr-strain and ammonia adsorbed graphene (in all the configurations), except $\sigma^*$ bands. The shift is more in the case of D configuration than that of U-type adsorption. Band gap opening is found to be nearly the same for both graphene under strain and graphene/NH$_3$ system. Essentially, the band gap is mainly contributed by the strain developed on graphene due to adsorption of ammonia, not due to the charge transfer.

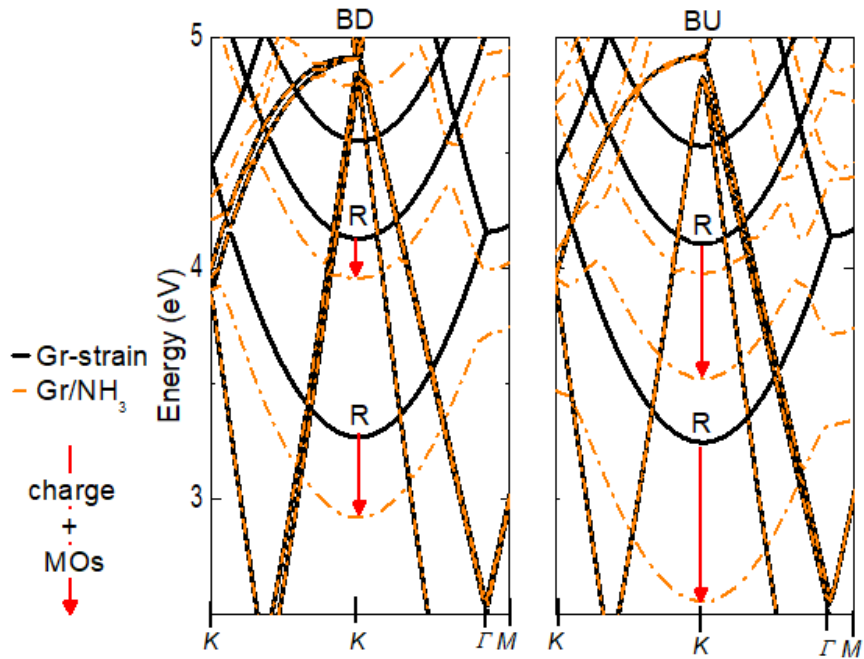

**SFIG 2: Electronic band structure of graphene/NH$_3$ system for BD and BU configurations. R indicates the conduction band replicas due to strain. Red arrow shows the shift of the replicas towards EF due to the presence of MOs of NH$_3$ and any associated charge transfer.**